\documentclass[10pt]{article}

\usepackage{amsmath,epsfig}
\usepackage{amssymb}

  \topmargin
11in \advance \topmargin -\textheight \divide \topmargin by 2
\advance \topmargin -1.5in \leftmargin 8.5in \advance \leftmargin
-\textwidth \divide \leftmargin by 2 \advance \leftmargin -1in
\oddsidemargin \leftmargin \evensidemargin \leftmargin

\def\QED{\QEDclosed} % default to closed

\begin{document}

\title{A Practical Approach to Joint Network-Source Coding}
\author{Nima Sarshar and Xiaolin Wu \\
Department of Electrical and Computer Engineering \\
McMaster University, Hamilton, Ontario, Canada \\
{\small sarshan@mcmaster.ca/xwu@ece.mcmaster.ca} }

\maketitle

\begin{abstract}
We are interested in how to best communicate a real valued source to
a number of destinations (sinks) over a network with capacity
constraints in a collective fidelity metric over all the sinks, a
problem which we call joint network-source coding.  It is
demonstrated that multiple description codes along with proper
diversity routing provide a powerful solution to joint
network-source coding. A systematic optimization approach is
proposed. It consists of optimizing the network routing given a
multiple description code and designing optimal multiple description
code for the corresponding optimized routes.

\end{abstract}
\section{Introduction}
\subsection{Joint Network-Source Coding: Problem
Formulation}\label{sec:form}

Joint network-source coding (JNSC) is the problem of communicating
and reconstructing a (usually real valued) source in a network to
a maximal collective fidelity over a given set of sinks, while the
flows of the code streams satisfy the edge capacities of the
network.  JNSC can be considered as a lossy version of the
(lossless) network coding problem, since the reconstruction is not
necessarily perfect. The source is "observed" by a subset of nodes
in the network, called source nodes. Due to capacity constraints,
source nodes have to communicate a coded version of the source to
their neighboring nodes.  Just as in lossless networked coding,
intermediate nodes can in general transcode data received from
other nodes, and communicate it to their neighbors. Any node in
the network, based on the information it receives about the
source, can reconstruct the source with some distortion.

Unlike its lossless counterpart, the interaction of lossy
source-network codes with arbitrary networks is largely
unexplored.
%While recent years has seen a tremendous progress in
%lossless networked coding theory and techniques, lossy networked
%communication has received little, if any, attention.
In fact, the term network coding refers, almost exclusively, to
lossless network communication. This is despite the fact that
arguably the majority of
%resource intensive
applications, both in the Internet and in various wireless setups,
involve lossy source communication, in particular for multimedia
applications. It should however be noted that some of the well
studied examples of multi-terminal   source coding problems (e.g.,
multiple-description coding) are simple examples of a general
lossy networked coding problem.

In this paper the network model is similar to, now standard,
models in network coding \cite{ash}.  The JNSC problem is defined
by the following elements: \\ (1) A directed graph $G\langle V,E
\rangle$.\\ (2) A source $X$ in some alphabet $\Gamma$ and a set
of distortion measures $\rho^n:\Gamma^n \rightarrow \mathbb{R}^+$.
We assume $X$ admits a rate-distortion function $D_X(.)$, with
$\rho$ as the measure.\\ (3) A function
$R:E\rightarrow\mathbb{R}^+$ that assigns a capacity $R(e)$ to
each link $e\in E$. We normalize bandwidth with the source
bandwidth, therefore, $R(e)$ is expressed in units of bits per
source symbol. \\   (4) Two sets $S,T \subseteq V$ that denote the
set of source and sink nodes respectively.  The source nodes
observe, encode, and communicate $X$ in the network. Source nodes
are assumed to be able to collaborate in encoding. This can model,
for example, computer networks where sources are encoded off-line
and copies of the code are distributed to the source nodes.

Nodes can communicate with neighbor nodes at a rate bounded by the
capacity of the corresponding link. The goal is to communicate the
source $X$ from the source nodes in $S$, and reconstruct $X$ at
the sink nodes in $T$. A distortion vector $\mathbf{d}=(d_t,t\in
T)\in \mathbb{R}^{|T|}$ is said to be achievable if $X$ can be
reconstructed with a maximum distortion of $d_t$ at a sink node
$t$ by using a coding scheme that respects the capacity
constraints on the links, i.e., the rate of information per source
sample communicated over $e$ is less than $R(e)$.  As in
\cite{ash}, we need to leave the details of the code unspecified,
because it proves extremely hard to come up with the most general
class of possible codes. An intriguing problem is how to
characterize the set of all achievable distortion $t$-tuples
${\cal D}_X(G,S,T,R) \subset \mathbb{R}^{|T|}$.   In this paper,
on the other hand,
 we are interested in a  coding schemes that minimizes a
weighted average distortion over all the sink nodes, that is,
$\sum_{t\in T} p_t d_t$ for some weighting vector
$\mathbf{p}=(p_t; t\in T)$.

\subsection{Multiple-Description Coding: a Tool for JNSC}

Multiple-description codes (MDC) have always been associated with
robust networked communications, because they are designed to
exploit the path and server diversities of a network.  The present
active research on MDC is driven by growing demands for real-time
multimedia communications over packet-switched lossy networks,
like the Internet.  With MDC, a source signal is encoded into a
number of code streams called descriptions, and transmitted from
one or more source nodes to one or more destinations in a network.
An approximation to the source can be reconstructed from any
subset of these descriptions.
%Ideally, descriptions are communicated
%through different paths to take advantage of various sources of
%diversity in the network.
If some of the descriptions are lost,
%due to the lossy nature of the network,
the source can still be approximated by those received. This is
why there seems to be a form of consensus in the literature in
that multiple description codes should only be used in
applications involving packet loss, because only in this case the
overhead in the communication volume can be justified.

This paper shows, however, that MDC is beneficial for lossy
communication even in networks where all communication links are
error free with no packet loss.  In this case, multiple
description coding, aided by optimized routing, can improve the
overall rate-distortion performance by exploiting various paths to
different nodes in the network.  This is best illustrated through
some examples.
\subsubsection{Example 1}
\begin{figure}
\begin{center}
\includegraphics[width=2.5in,height=1.5in]{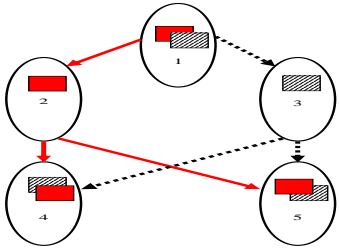}
\caption{An example of flow of a two description
code.}\label{fig:1}
\end{center}\vspace{-.3in}
\end{figure}

In Fig.\ \ref{fig:1}, a source node (node 1) feeds a coded source
into a network of four sink nodes (nodes 2-5). The goal is to have
the best reconstruction of the source at each of these four nodes.
All link capacities are $C$ bits per source symbol. MDC encodes
the source into two descriptions (shown by solid and dashed boxes
in the figure), each of rate $C$. Descriptions 1 and 2 are sent to
nodes 2 and 3 respectively. Node 2 in turn sends a copy of
description 1 to nodes 4 and 5, while node 3 also sends a copy of
description 2 to nodes 4 and 5.  In the end, nodes 4 and 5 will
each receive both descriptions, while nodes 2 and 3 will only
receive one description.

To see how the nodes in the network benefit from MDC, let
$D_1(C),D_2(C),D_{12}(C)$ be the distortion in reconstructing the
source given description 1 or 2 or both.  Let
%$\mathbf{\delta}=(\delta_2,\delta_3,\delta_4,\delta_5)$
$\mathbf{d}=(d_2,d_3,d_4,d_5)$ be the vector of the average
distortions in reconstructing the source at nodes 2 through 5.
Therefore, $\mathbf{d}=(D_1(C),D_2(C),D_{12}(C),D_{12}(C))$.

Let's define $\mathcal{D}_M$ as the set of all achievable
distortion 4-tuples $\mathbf{d}$.  Although MDC is in general a
special form of lossy networked coding, $\mathcal{D}_M$ still
contains a large and interesting subset of all achievable
distortion tuples. In this example, it includes for instance, the
distortion region achievable by separate source and networked
 coding. By results in \cite{ash}, the maximum rate
with which common information can be communicated to nodes 2
through 5 is $C$ bits per source symbol. Therefore, the distortion
rate achievable by separate source and network coding is
$\mathcal{D}_S =\{(\delta_2,\delta_2,\delta_4,\delta_5):
\delta_i\geq 2^{-2C}, i=2,3,4,5\}$.  We immediately have that
$\mathcal{D}_S \subset \mathcal{D}_M$ by noting that
$D_1(C)=D_2(C)=D_{12}(C)=2^{-2C}$ is part of $\mathcal{D}_M$. In
fact this corresponds to communicating two identical descriptions,
each of which is an optimal (in the rate-distortion sense) source
code of rate $C$ for $X$.

The inefficiency of separate source and network coding lies in
that even though nodes $4,5$ have twice the incoming capacity
compared to nodes $2,3$, their reconstruction error ($d_4=d_5$) is
bounded by the reconstruction error of the weaker nodes
($d_2=d_3$).  Unlike lossless coding, lossy codes can play a
tradeoff between the reconstruction errors at different nodes,
generating a much larger set of achievable distortion tuples
$\mathbf{d}$ than $\mathcal{D}_S$. These tradeoffs are essential
in practice. For instance, in networked multimedia applications
over the Internet, where the network consists of a set of
heterogenous nodes, the experience of a user with broadband
connection should not be bounded by that of a user with a lesser
bandwidth. Such tradeoffs are perhaps best treated as an
optimization problem by introducing appropriate Lagrangian
multipliers (or weighting functions). An objective function to
minimize, therefore, can be defined as
\begin{equation}\label{eqn:ovd}
\overline{d}(\mathbf{p},\mathbf{d})=\mathbf{p}^T \cdot \mathbf{d}
\end{equation}
where $\mathbf{p}=(p_2,p_3,p_4,p_5)$ is an appropriate weighting
vector. An optimal solution will be given by:
\[
\overline{d}^*(\mathbf{p})=\min_{\mathbf{d} \in \mathcal{D}_M}
\overline{d}(\mathbf{p}, \mathbf{d})
\]

Once the optimal distortion vector $\mathbf{d}^*$ is found, one
should, in principle, be able to find a multiple description code
that provides the marginal and joint distortions corresponding to
$\overline{d}^*(\mathbf{p})$ (such an MDC exists).

As a concrete example, let's optimize the average distortion at
all nodes 2 through 5 in Fig. \ref{fig:1} for
$\mathbf{p}=(1/4,1/4,1/4,1/4)$, in which case:
\begin{equation}\label{eqn:1}
\overline{d}=\frac{2D_{12}(C)+D_1(C)+D_2(C)}{4}
\end{equation}
To be specific, lets assume that the source in question is an iid
Gaussian with variance one for which achiveable distortions in
multiple description coding are completely derived by Ozarow in
\cite{oz}. The symmetry in indices 1 and 2 ensures that
(\ref{eqn:1}) is minimized when the two descriptions are balanced,
that is, $D_1(C)=D_2(C)=D$.  Ozarow's result, when specialized to
balanced MDC states that the following set of distortions are
achievable:
\begin{eqnarray}\label{eqn:2}
D_1 &=& D_2 = D \geq 2^{-2C} \\
D_{12} &\geq&
\frac{2^{-4C}}{(D+\sqrt{D^2-2^{-4C}})(2-D-\sqrt{D^2-2^{-4C}})}\nonumber
\end{eqnarray}

\begin{figure}
\begin{center}
\includegraphics[width=2.5in,height=2.0in]{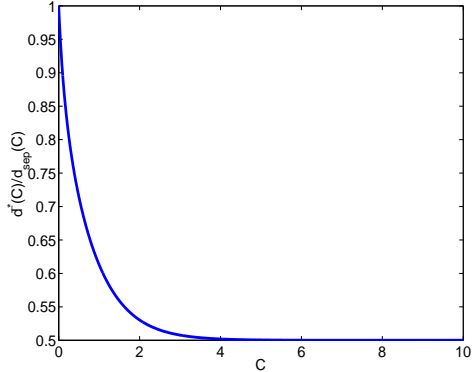}
\caption{The ratio of minimum average distortion obtained by an
optimized MDC code over the one obtained by separate source and
network coding, for the example in Fig. \ref{fig:1}.}\label{fig:2}
\end{center}\vspace{-.2in}
\end{figure}

The average distortion in (\ref{eqn:1}) can therefore be minimized
under the constraints of (\ref{eqn:2}). This is a particularly
easy task because the region (\ref{eqn:2}) is convex. Let this
optimal average distortion be $\overline{d}_M^*(C)$. By separating
source from network coding, the reconstruction distortion at nodes
2 through 5 (and hence the average distortion over all these
nodes) is at best $d_S(C)=2^{-2C}$.  It is easy to show that
$\overline{d}_M^*(C) < d_S(C)$ for all $C>0$ (see Fig.
\ref{fig:2}). In other words, for all $C>0$ there exists a
balanced two description code for which the average distortion
over all sink nodes is strictly less than the average distortion
achievable by any separate source and network  coding scheme.

\subsubsection{Example 2}

\begin{figure}
\begin{center}
\vspace{-1in}\includegraphics[width=4.5in,height=4.5in]{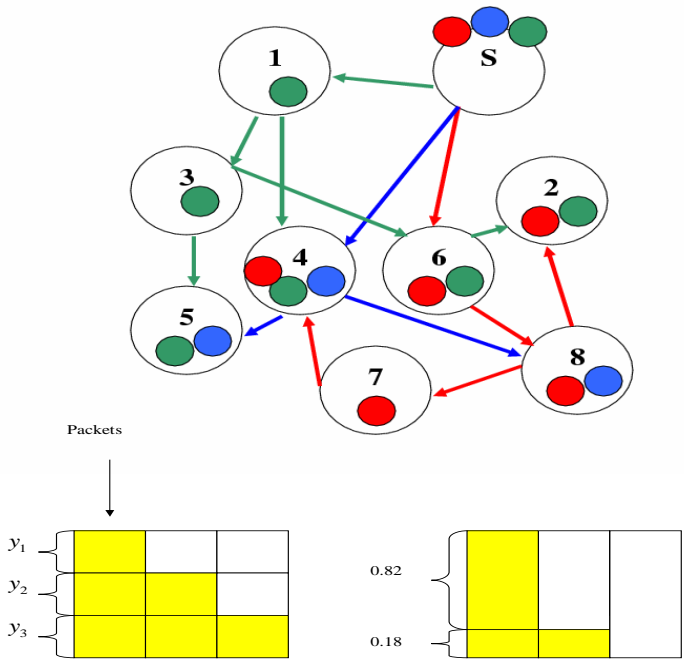}
\caption{An example of optimal flow of a three description code.
The bottom of the figure depicts the procedure of designing a
3-description MDC using Priority Encoding Transmission technique
as explained in Section \ref{sec:design}.}\label{fig:exp}
\end{center}
\end{figure}

As a perhaps more involved example, consider the problem depicted in
Fig. \ref{fig:exp}. A source at node $S$ is to be communicated to
nodes $1-8$. The goal again is to minimize the average distortion
over all these 8 sink nodes. All links have capacity $C$ bits per
source symbol. We choose to use an MDC with $3$ descriptions each of
rate $C$. These descriptions are indicated by three colors, Red,
Green and Blue. Fig. \ref{fig:exp} shows a routing strategy that
delivers these descriptions optimally to all the 8 nodes. The
routing is optimal since each node receives a number of distinct
descriptions exactly equal to its incoming capacity.

Once the routing is optimized, one still needs to optimize the MDC.
Unlike the case of two descriptions, there is no closed form
representation of the rate-distortion behavior of a balanced
3-description code even in the case of a Gaussian source. Therefore,
we resort to the practical technique of Priority Encoding
Transmission (PET) for producing whatever required number of
balanced descriptions.  The class of PET codes
%The rate distortion behavior of the codes designed with PET are
are easily parameterizable.

In particular, as is shown in Section \ref{sec:design}, for any
progressively refinable source with distortion-rate function
$D(R)$, an MDC of $K$ descriptions each of rate $r$ per packet can
be constructed using PET such that the distortion given any $k\leq
K$ of the descriptions, the source can be reconstructed with
distortion at most $D(r\sum_{l=1}^k l)$, where $\mathbf{y}=(y_l;
l=1,2,...,K)$ is any positive vector such that $\sum_{l=1}^K
y_l=1$. There is a one-to-one correspondence between an MDC code
and any such vector $\mathbf{y}$. Optimizing the MDC within the
class of PET codes will result into a very convenient convex
optimization problem. For the example in Fig.~\ref{fig:exp},
three, four and one nodes receive 1,2,3 distinct descriptions
respectively. For a PET code with vector
$\mathbf{y}=(y_1,y_2,y_3)$ the average distortion at the sinks,
assuming a Gaussian source of variance one, can be written as:
\[
\overline{d}=8^{-1}\left(3\cdot 2^{C y_1}+4\cdot 2^{C
(y_1+2y_2+3y_3)}+ 2^{C (y_1+2y_2+3y_3)} \right)
\]
Minimizing the above over all $y_1,y_2,y_3 \geq 0$ such that
$y_1+y_2+y_3=1$ is a convex optimization problem with linear
constraints. For $C=1$, the optimal solution is $y_1=0.82, y_2=0.18,
y_3=0$.

\subsection{Design Issues}

Above examples make a number of important observations and expose
some design issues, which are the subject of the rest of this paper.
\begin{itemize}
\item MDC routing can exploit path diversity in ways that a
separate source and network coding can not. For instance, in the
example of Fig.~\ref{fig:1}, nodes 4 and 5 can benefit from the data
received both from nodes 2 and 3, while nodes 2 and 3 themselves can
benefit from the data they relay, which was not possible if a common
data was communicated to both nodes 2 and 3 by the source node.\item
To benefit from MDC in the network, routing needs to be
optimized.

\item Not only the routing, but also the MDC should be
designed optimally. Our strategy will be to optimize MDC generated
through PET.

\item Although in the above two examples, optimizing the routing
of MDC descriptions and optimizing the MDC code happen to be
separable, the two optimizations in general, should be carried out
jointly. Due to complexity concerns, our strategy is to carry out
the optimizations separately.

\item Unlike the two examples reported in this section,
optimizing the MDC may result in a different number of, potentially
unbalanced, descriptions. In this paper, we confine ourselves solely
to the case of balanced MDC of the same rate. The \emph{total number
of descriptions and their rates} however, are left as optimization
parameters.
\end{itemize}

In the next section, we consider the problem of optimal diversity
routing of MDC packets, called Rainbow Network Flow (RNF) problem.
MDC codes designed through PET is introduced in Section
\ref{sec:design} where we discuss  JNSC through optimized
diversity flow and optimal MDC design. Section \ref{sec:sim}
examines the issues regarding the choice of description rates and
the total number of descriptions, and presents simulation results
for a family of randomly generated directed acyclic network
graphs.

\section{Rainbow Network Flow Problem}\label{sec:rnf}
Rainbow Network Flow (RNF) introduced in \cite{ISIT05,NetCod} is the
problem of optimal routing of MDC packets in a general network. RNF
is different from usual commodity flow in that one should take into
account the information content (or color) of the descriptions. In
particular, receiving duplicate descriptions is not beneficial in
reconstructing the source. Also, unlike commodity flow, information
packets can be duplicated at intermediate nodes. A node desires the
\emph{rainbow} effect by having as many distinct descriptions (or
colors) as possible.

%RNF problem receives a choice of MDC as its input and will find
%the optimal routing of its descriptions to minimize the average
%distortion at the sink nodes. We now state a version of RNF
%problem most appropriate to our discussion.

The RNF problem is defined with following inputs: \\
(1) $G\langle V,E\rangle $, a directed
graph with a node set $V$
and an edge set $E$.\\
(2) $S=\{s_1,s_2,...,s_{|S|}\},T=\{t_1,t_2,t_3,...,t_{|T|}\}$ two
subsets of $V$ representing the set of source and sink nodes
respectively.\\
(3)  A function $R:E\rightarrow \mathbb{R}^+$
representing the capacity of each link in $G$.\\
(4) A set $\chi\subset \mathbb{N}$ called the description set.\\
(5) An $r\in \mathbb{R}^+$ called the description rate.\\
(6) $\delta:\{1,2,...,|\chi|\}\rightarrow \mathbb{R}^+$ ,
a non-increasing function specifying the choice of balanced MDC.
$\delta(k)$ is the reconstruction distortion when
any subset of size $k$ out of $K=|\chi|$ possible descriptions
are present at the decoder. \\
(7) $\mathbf{p}=(p_t; t\in T)$ a positive vector that
weighs the importance of each sink node $t\in T$. \\

A special and important form of the RNF problem is when
$\delta({\cal M})=1-|{\cal M}|/|\chi|$ for any ${\cal M}\subset
\chi$. In this case, the distortion is simply given by the size of
the subset of available descriptions. Since $\delta({\cal M})$
depends only on the cardinality of the set ${\cal M}$, this
particular RNF problem is called cardinality RNF (CRNF).

The goal of RNF problem is to find routing (or flow) paths that
take descriptions from source nodes to sinks in a way that
minimizes a weighted average distortion at the sink nodes.

A flow path from $s\in S$ to $t\in T$ is a sequence of edges
%$w(s,t)=\left((v_0=s,v_1),(v_1,v_2),...,(v_{m-1},v_{m}=t)\right)$,
$w(s,t)=[(v_0=s,v_1),(v_1,v_2)$,$\cdots$,$(v_{m-1},v_{m}=t)]$, such
that $(v_i,v_{i+1})\in E$ for $i=0,1,...,m-1$.

A rainbow network flow, denoted by $\alpha(W,f)$, consists of a set
$W$ of flow paths in $G$,
%$W\subset {\cal W}(G,S,T)$
and a so-called flow coloring function $f: W \rightarrow \chi$ that
assigns a description (or color) in $\chi$ to each flow path. For
the flows in Fig.\ \ref{fig:1}, $W=\{ [(1,2),(2,4)]$,
$[(1,2),(2,5)]$, $[(1,3),(3,4)]$, $[(1,3),(3,5)] \}$, and
%all the possible?
%four colored flow paths:
%($W={\cal W}(G,S,T)$).
the flow coloring function $f$ assigns $f([(1,2),(2,4)])=1$,
$f([(1,2),(2,5)])=1$, $f([(1,3),(3,4)])=2$, $f([(1,3),(3,5)])=2$.

Let $\Phi_E(e,W)$ and $\Phi_V(v,W)$ be the sets of all colored
flow paths in $W$ that contain the link $e$ or the node $v$,
respectively.  For example,
$\Phi_E(e=(1,2),W)=\{[(1,2),(2,3)],[(1,2),(2,5)]\}$.

The spectrum of an edge $e\in E$, with respect to RNF $\alpha$, is
defined as:
\[
\Psi_E(\alpha,e) \equiv \bigcup_{w \in \Phi_E (e,W)} f(w)
\]
Likewise, the spectrum of a node $v$ is defined as:
\[
\Psi_V(\alpha,v) \equiv \bigcup_{w\in \Phi_V (v,W)} f(w)
\]

In Fig.\ \ref{fig:1} for instance
$\Psi_E((1,2))=\Psi_E((2,4))=\Psi_E((2,5))=\{1\}$ and,
$\Psi_E((1,3))=\Psi_E((3,4))=\Psi_E((3,5))=\{2\}$.  The spectrum
of the nodes $4,5$ consists of both descriptions (i.e.,
$\{1,2\}$), while the spectrum of the nodes $2,3$ is $\{1\},\{2\}$
respectively.

A rainbow network flow $\alpha(W,f)$ is said to be admissible with
capacity function $R$, if and only if:
\begin{equation}\label{eqn:cap}
|\Psi_E(\alpha,e)|<R(e) \qquad \forall e\in E
\end{equation}
The significance of this inequality is that it allows the
duplication of a description by relay nodes. Therefore, two flow
paths of the same color can pass through a link $e$, and yet
consume a bandwidth of only $r$. The rainbow flow plotted in
Fig.~\ref{fig:1} is admissible because at most one description
with rate $C$ is communicated over each link and the capacity of
each link is $C$. This is made possible by duplicating at nodes
2,3.

For a given $\alpha(W,f)$, $|\Psi_V(\alpha,v)|$ is the number of
distinct descriptions (out of a total of $K=|\chi|$ such
descriptions) available to node $v$. If the node $v$ is a sink
node, the reconstruction distortion at $v$ will therefore be:
\begin{equation}\label{eqn:dis_v}
d_t= \delta(|\Psi_V(\alpha,t)|)
\end{equation}

The weighted average distortion at all the sink nodes is then:
\begin{equation}\label{eqn:avg}
\overline{d}(\alpha)= |T|^{-1} \sum_{t \in T}p_t
\delta(|\Psi_V(\alpha,t)|)
\end{equation}

RNF problem is therefore, that of finding an admissible rainbow
network flow $\alpha^*$ that minimizes (\ref{eqn:avg}).

Unfortunately, RNF problem was proved to be NP-hard in its full
generality \cite{ISIT05, NetCod}. It is however polynomially
solvable for some practically important cases (e.g., for tree
network topologies). For general directed acyclic graphs (DAG), the
problem has an amenable integer-convex programming formulation. In
the next section, we use such a formulation for finding solutions to
relatively large instances of up to 1000 nodes.

%An optimized solution to RNF problem with proper MDC design will
%provide

\section{Code Design}\label{sec:design}

%RNF problem  receives the choice of the MDC as its input. Function
%$\delta$, the set $\chi$ and description rate $r$ are all inputs to
%the problem.

An optimal solution to the JSNC problem requires joint optimization
of the MDC code and rainbow network flow.  This, however, is an
extremely hard task, given that most versions of the RNF problem are
intractable.  In quest for a practical solution, we will resort to
iterative and approximate numerical methods.  We use a certain
family of balanced multiple description codes that are completely
parameterizable. This allows us to formulate the code design for a
fixed rainbow network flow as a convex optimization problem.

\subsection{MDC using PET}
%For producing balanced MDC, we use a popular method called Priority
%Encoding Transmission (PET),

The PET technique can produce any number of balanced multiple
descriptions out of a progressively encoded source stream. The
idea is the following. To make $K$ balanced descriptions each of
rate $r$ bits per source symbol, for a large enough value of $n$,
encode $n$ samples of $X$ into a progressive bitstream
$(b_0,b_1,...,b_{nrL})$, where we assume $n\cdot r$ is an integer
for simplicity. Take a $K\times (n\cdot r)$ binary matrix and call
it $Y=[Y_{ij},i=1,2,...,K,j=1,2,...,n\cdot r]$. Now take
$\mathbf{y}=(y_i,i=1,2,...,K)$, any vector of real numbers of
length $K$ such that $\sum_{i=1}^K y_i=1$.  For $i=1,2,...,K$ do
the following: let $Y_l,Y'_l$ for $l=1,2,...,K$ be sub-matrices of
$Y$ consisting of:
\begin{eqnarray*}Y_l=[Y_{ij}&;&\: i=1:l,\quad
j=\sum_{k=1}^{l-1} n\cdot r y_k:\sum_{k=1}^{l} n\cdot r y_k]\\
Y'_l=[Y'_{ij}&;&\: i=l+1:K,\quad j=\sum_{k=1}^{l-1} n\cdot r
y_k:\sum_{k=1}^{l} n\cdot r y_k]
\end{eqnarray*}

Therefore, matrix $Y_i$ contains $i\cdot n \cdot r\times y_i$ bits
while $Y'_i$ has $(K-i)\cdot n\cdot r\times y_i$ bits. For
$i=1,2,...,K$, put the $i\cdot n \cdot r\cdot y_i$ bits of the
progressive source code stream , from $b_{g(i)}$ to
$b_{g(i)+i\cdot n\cdot r\cdot y_i}$ in $Y_i$, where
$g(i)=\sum_{k=1}^i k\cdot n\cdot r y_k$. In $Y'_i$ on the other
hand, put parity symbols of a $(i\cdot n\cdot r\cdot y_i,K\cdot
n\cdot r\cdot y_i )$ ideal \textit{erasure correction} code
corresponding to the bits in $Y_i$.

Now the descriptions consist of the $K$ columns of the matrix $Y$,
each of $n\cdot r$ bits. The total source bits used is $n\cdot r
\sum_{k=1}^K k y_k$. It is easily verified that given any $l\leq
K$ descriptions, the first $\xi_l=\sum_{k=1}^l k\cdot n\cdot
r\cdot y_k$ bits of the source bitstream can be recovered. For
large enough $n$ and assuming the source is progressively
refinable, given any $k$ distinct descriptions, the source can
therefore be reconstructed within  distortion:
$D_X(\xi_k/n)=D_X\left(r \sum_{l=1}^k l y_l\right)$.

This is schematically depicted in bottom of Fig.~\ref{fig:exp} for
a 3-description code (K=3). "Yellow" bits indicate source bits
while "white" bits are parity symbols. The description "packets"
are the columns of this matrix. Vectors $\mathbf{y}$ therefore
parameterize the space of all MDCs that can be generated through
PET. It is over this space that we will carry out our code
optimization.

\subsection{Optimizing Code for a Fixed Rainbow Flow}
For any admissible flow $\alpha$, define the rainbow flow vector
(RFV), $\mathbf{q}(\alpha)=(q_t; t\in T)$ such that:
\[
q_t= |\Psi_V(\alpha,v)|
\]
In other words, $q_t$ is the number of distinct descriptions
available to a sink node $t$. Suppose a
%(perhaps suboptimal)
solution $\alpha^*$ is found to the RNF problem with respect to an
MDC, and the rainbow flow $\alpha^*$ produces an RFV
$\mathbf{q}^*$. Lets replace this MDC for which the original RNF
was optimized with the MDC designed through a PET technique.  Let
the latter MDC be specified by the vector $\mathbf{y}$.  Then, the
weighted average distortion in (\ref{eqn:avg}), now a function of
$\mathbf{q}^*$ and $\mathbf{y}$, can be written as:
\begin{equation}\label{eqn:opty}
\overline{d}(\mathbf{y},\mathbf{q}^*)=|T|^{-1} \sum_{t\in T} p_t
D_X \left(r \sum_{l=1}^{q_t^*} l y_l\right )
\end{equation}
An optimal MDC can be found as an answer to the following problem:
\begin{equation}\label{eqn:sol}
\min_{\mathbf{y}\succ 0,|{\mathbf y}|_1=1}\;\; \sum_{t\in T} p_t
D_X \left(r \sum_{l=1}^{q_t^*} l y_l\right )
\end{equation}

Since $D_X(\cdot)$ is a convex function, this is a convex
optimization problem with linear constraints and can be solved
efficiently using standard tools.

\subsection{An Optimized Solution to JSNC Problem}
We can now devise a systematic procedure for finding an optimized
solution to the Joint Source Network Coding problem. We approach
this by first solving the RNF problem.  We are particularly
interested in finding the solution $\alpha^*$ to CRNF, because the
optimization of the MDC flow is carried out regardless of the
particular choice of the MDC and the statistics of the underlying
source. As stated in Section \ref{sec:design}, the goal of CRNF is
to find a flow that maximizes the sum of distinct descriptions
received by all the sink nodes. This approach is most reasonable
when the reconstruction at all the sink nodes is equally important
(e.g., $p_t=1, \forall t\in T$).

Given an optimized RNF and the resulting RFV, the MDC is optimized
by solving (\ref{eqn:sol}).

\section{Numerical Simulations}\label{sec:sim}
The proposed JSNC approach is tested on a number of randomly
generated directed acyclic network graphs.  The simulation results
are examined in this section.

%where a discussion on the choice of the rate of descriptions $r$ as
%well as their total number $K$ is provided.

\subsection{0-1 Linear Integer Programming for CRNF}

For directed acyclic graphs (DAG), the CRNF can be posed as a 0-1
linear integer programming problem. Let $\chi=\{1,2,...,K\}$ be the
set of all descriptions. For each edge $e=(v,v') \in E$, define a
binary variable $x^k_{v,v'}$ that is 1 if $k \in \Psi_E(\alpha,e)$
and 0 otherwise. Furthermore, for each node $v$, we use a binary
variable $y^k_v$ to indicate whether $k \in \Psi_V(\alpha,v)$, that
is, whether description $k$ is received by $v$. For $v\in S$ that is
a source of description $k$, $y^k_v=1$ automatically. For $v$ that
is not a source of the color $k$, obviously
\[
y^k_v \leq \sum_{v'\in \ss(v)}x^k_{v,v'}
\]
where ${\ss}(v)$ is the set of all nodes that have a link into $v$.

A relay node can duplicate descriptions it receives from other
nodes, hence
\begin{equation}
\forall j \ \  x^k_{i,j} \leq y^k_i . \label{duplicate-constraint}
\end{equation}

The edge capacity constraint is:
\begin{equation}
\label{eq-directed} \forall e=(v,v')\in E, \ \  r \sum_{k}
x^k_{v,v'} \leq R(e)
\end{equation}

CRNF problem for DAG reduces to finding the binary variables
$x^k_{v,v'}$ that that maximizes the sum of the number of distinct
descriptions received by all the sink nodes, that is, $\sum_{t\in
T} y^k_t$. For a fixed number of descriptions $K$ and rate $r$,
this is a linear binary program with linear constraints. For
values of $K$ less than 10, our commercial package was able to
solve instances of CRNF problem of 1000 nodes in under one minute.

\subsection{Network Simulation Setup}

We produce a family of DAGs motivated by a simplistic model for
the growth of peer-to-peer networks. We start from a single node.
At each step, a new node is added. Then $m$ nodes are chosen at
random with replacement from the existing nodes, and a link is
made from each of the $m$ chosen nodes to the new node. Once the
network grows to $N$ nodes, we assign a capacity $C(e)$ to each
link $e$, where $C(e)$ is a random integer between 1 and $C_{max}$
for some choice of maximum capacity $C_{max}$.  For integer edge
capacities, the optimum solution to the CRNF problem will consist
of all integer flows. We use an iid Gaussian source of variance
one in all cases.
% (?? is that true??).

We take the rate of the descriptions to be $1$. It can also be
shown that for a fixed source and network, increasing the number
of descriptions $K$ will not increase the overall distortion but
the running time of the integer programming will increase
exponentially in $K$.  In this consideration, we always start with
the smallest possible $K$, and increase $K$ until the overall
distortion does not decrease any more.

\begin{figure}
\begin{center}
\includegraphics[width=2.5in,height=1.75in]{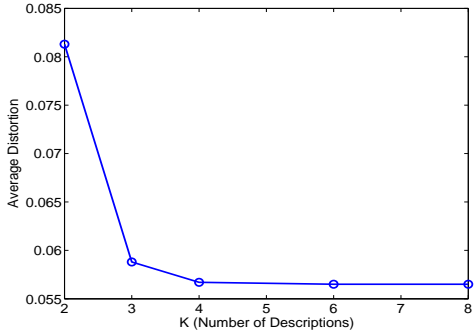}
\caption{Distortion as a function of the number of descriptions.
Description rate is $r=1$, the network has $N=50$ nodes and
$C_{max}=3$ and $m=3$.}\label{fig:con}
\end{center}
\end{figure}

For a network of size $N=50$, with nodes of in-degree $m=3$ and
$C_{max}=3$, the overall distortion $\overline{d}$ is optimized by
first solving the instance of CRNF problem and then optimizing the
MDC. The optimization is repeated for increasing number of
descriptions $K$. The distortion converges to its final value for
$K=6$ as shown in Fig. \ref{fig:con}. The overall optimization
took less than 5 seconds.

\begin{figure}
\begin{center}
\includegraphics[width=2.5in,height=1.7in]{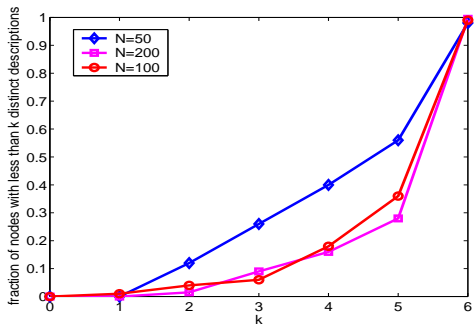}
\caption{The fraction of nodes that receive less than or equal to
$k$ distinct descriptions as a function of $k$ for different
network sizes. We have $C_{max}=3$, $m=3$ and $K=6$. Therefore,
$k=0,1,...,6$. For larger network sizes, due to an increase in the
number of available paths, the fraction of nodes that receive
higher number of distinct descriptions increases.}\label{fig:inc}
\end{center}
\end{figure}

To see the effect of enlarging the network, we have provided
simulation results for fixed $K=6$, $C_{max}=3$ and $m=3$ and
different networks sizes. For the above family of random networks,
the increase in the network size will offer greater path diversity
to the more recently created nodes which are at the bottom of the
network hierarchy.  Therefore, as the network size grows, the
fraction of nodes that receive higher number of descriptions
increases (see Fig.\ \ref{fig:inc}), which leads to a decrease in
the overall average distortion.

\section{Conclusion}
We introduced the problem of joint network-source coding (JNSC) in
which the goal is to best communicate a real valued source to a
number of destinations using the collaboration of all nodes in the
network. We found that multiple description coding is a powerful
tool for exploiting path diversity in a network.  We provided a
systematic approach for optimizing the routing of descriptions as
well as optimally designing the MDC. To our best knowledge, this
is the only known formulation and systematic approach to joint
network-source coding problem.

Our ongoing research is on an iterative optimization approach in
which, (1) optimize the flow of descriptions for a given MDC and,
(2) optimize in turn the MDC with respect to the resulting flow,
(3) continue the process iteratively until there is no further
reduction in the average distortion.

We are also applying our approach to a peer-to-peer networking
scenario in which the goal is to have a real-time presentation of
a multimedia content in the network.  In this case, maximizing the
description diversity has to be done locally at individual nodes,
and the network dynamics should also be taken into account.

\begin{table}
\center
\begin{tabular}{|c|c||c|c|c|c|c|c|}
  \hline
   N & $\overline{d}^*$ & $y_1$ & $y_2$ & $y_3$ & $y_4$ & $y_5$ &
   $y_6$ \\
   \hline
      \hline
   50 & 0.1104& 0.0 & 0.7069 & 0.2067 & 0.083 & 0.0 & 0.0\\
   \hline
    100 & 0.0466& 0.0 & 0.445 & 0.070 & 0.396 & 0.090 & 0.0\\
   \hline
   200 & 0.0302& 0.0 & 0.204 & 0.556 & 0.094 & 0.117 & 0.030\\
   \hline
\end{tabular}\caption{The optimal vector $\mathbf{y}$ for
different network sizes. $C_{max}=3$, $K=6$ and $m=3$.}
\end{table}

\vspace{-.1in}
\bibliographystyle{abbrv}

\end{document}